\documentstyle[preprint,aps]{revtex}
 
 \def\beq{\begin{equation}}
 \def\eeq{\end{equation}}
 \def\beqa{\begin{eqnarray}}
 \def\eeqa{\end{eqnarray}}
 \def\nn{\nonumber}
 \begin{document}
 \begin{titlepage}
 \begin{flushright}
 IC/97/56\\
 SISSAREF-/97/EP/80
 \end{flushright}
 \begin{center}
 {\Huge Interaction of moving D-branes on orbifolds}
 
 \vspace*{1cm}
 
 {\large  Faheem Hussain$^1$, 
 Roberto Iengo$^2$, Carmen N\'u\~nez$^3$ and Claudio A.
 Scrucca$^2$}
 \vspace*{0.5cm}
 
 $1$ International Centre for Theoretical Physics, Trieste, Italy\\
 $^2$ International School for Advanced Studies and INFN Sezione di 
 Trieste, Trieste, Italy\\
 $^3$ Instituto de Astronom\'{\i}a y F\'{\i}sica del Espacio (CONICET), 
 Buenos Aires,
 Argentina
 
 \vspace*{1cm}
 
 \end{center} 
 
 \begin{abstract}
We use the boundary state formalism to study the interaction of two moving
identical D-branes in the Type II superstring theory compactified on
orbifolds.
 By computing the  velocity dependence of the amplitude
in the limit of large separation we can identify the nature of the different
forces between the branes. In particular, in the $Z_3$ orbifold case we find
a boundary state which is coupled only to the $N=2$
graviton multiplet containing just a graviton and a vector like in the
extremal Reissner-Nordstr\"{o}m configuration. We also discuss other cases
including $T_4/Z_2$.    
 
  \end{abstract}
 
 \begin{flushleft}
 PACS: 11.25.-w\\
 Keywords: string theory, D-branes
 \end{flushleft}
 \end{titlepage}
 
 The non-relativistic dynamics of Dirichlet 
 branes \cite{bachas,lifschytz,porrati}, plays an essential role in the
 understanding 
 of string theory at scales shorter than the Planck length 
 \cite{kp,shenker}. It also provides some evidence \cite{lifschytz}  for 
 the existence of an underlying eleven dimensional theory \cite{mth}.
D-brane configurations have also been used to calculate the 
Bekenstein-Hawking formula for black hole entropy from a microscopic point
of view \cite{bh}.
  
In this work we use the boundary state technique \cite{pol2} to compute the
  interaction between two 
  moving identical D-branes in a TypeIIA or IIB superstring theory 
 compactified down to four dimensions on orbifolds.
We then evaluate these amplitudes in the limit of large separation of the
branes, that is in the field theory limit. Further by looking at the
dependence of these amplitudes on the rapidity we can unambiguously
identify the various contributions coming from the exchange of the
graviton, vector and scalar fields. The cancellation of the force at zero
velocity  for the toroidal  
compactification comes from the exchange of the $N=8$ graviton multiplet
containing the graviton, the
vectors and the scalars. In the case of the $Z_{3}$ orbifold there is a
particular BPS boundary state, corresponding to a D3-brane in TypeIIB,   
for which the no force
condition comes from the cancellation due to the exchange of the $N=2$
graviton multiplet containing the graviton and the graviphoton.
Thus, the classical solution corresponding to this configuration would be the
Reissner-Nordstr\"{o}m black hole. When
 the D-branes are on the fixed point of the orbifold, this being possible
for D0-branes in TypeIIA,  there are contributions from both the untwisted
and twisted sectors, the latter corresponding to the exchange of
vector multiplets containing vectors and scalars. 
For the case of the orbifold $T_4/Z_2\times T_2$
we find for parallel branes that the BPS
cancellation of the force is always like in the toroidal case, 
apart from when the branes are on fixed points where also 
additional vector multiplets contribute.
 
 We begin by considering a system of two D-branes in a
superstring theory 
 compactified down to four dimensions in the interesting case of the $Z_3$
orbifold, which breaks the 
supersymmetry down to N=2 (the branes will further break it to N=1)
\cite{min,hin}.
In the closed string picture the interaction between two branes is
viewed as the exchange of a closed string between two boundary states,
geometrically describing a cylinder. In the present work we
use $\tau$ for the coordinate along the length of the cylinder, $0\leq
\tau\leq l$, 
and
$\sigma$ as the periodic coordinate running from 0 to 1. 
We will always consider particle-like D-branes, that is 
 the time coordinate satisfies
  Neumann boundary conditions, $\partial_\tau X^0(\tau=0,\sigma)=\partial_\tau
 X^0(\tau=l,\sigma)=0$,
  whereas the three uncompactified space coordinates satisfy Dirichlet
 boundary conditions, $X^i(\tau=0,~\sigma)=0, X^i(\tau=l,\sigma)=Y^i$. 
The boundary conditions are implemented by suitable boundary states.

To evaluate the interaction between moving branes we calculate the 
amplitude 

\beq
{\cal A}=\int_{0}^{\infty}dl
<B,V,X^{i}=Y^{i}|e^{-lH}|B,V=0,X^{i}=0>\,,\label{ampl}
\eeq
where we have taken one of the branes to be at rest whereas the other is
moving with velocity $V$. The boundary states in position space are given in
terms of the momentum states as
\beqa
|B,V=0,X^{i}=0>&=&\int\frac{d^{3}k}{(2\pi)^{3}}e^{ik\cdot
(X=0)}|B,V=0,k>\nonumber\\
|B,V,X^{i}=Y^{i}>&=&\int\frac{d^{3}q}{(2\pi)^{3}}e^{iq\cdot Y}|B,V,q>\,.
\eeqa
In principle these should contain sums over the discrete compactified
momenta $p_{n}$ (they are zero for Neumann boundary conditions in
the corresponding directions whereas for Dirichlet compactified
directions
one has a wave function like for the uncompactified case). Since later
we will be interested in the field theory limit, $l\rightarrow \infty$,
and since each internal momentum $p_{n}$ will be weighted by a factor
$e^{-lp_{n}^{2}}$ from the compact Hamiltonian, only the $p_{n}=0$ term
of the sum will be relevant (similarly we ignore the winding states). 
Thus, the momentum content of the boundary state at rest will be
effectively $(k^0=0,\vec{k},p_n=0)$. 

We consider the case of two branes, one of them moving along one
 of the uncompactified space directions, say the $X^{1}$
direction. We make pairs of fields, $X^A=X^{0}+X^{1}$ and  
$X^B=X^{0}-X^{1}$, and pair the $X^{2}$ and $X^{3}$ into the complex
 fields $X^{2}\pm iX^{3}$.  The compact directions are
treated in terms of three pairs of complex cordinates \cite{min}.
 The net effect is as if the $b-c$ ghosts cancel 
 the contribution of the pair of coordinates, $X^{2}$ and $X^{3}$, orthogonal
to the boost. Similarly the $\beta-\gamma$ ghost
contribution cancels the contribution from the fermionic pair $\psi^{3}$
and $\psi^{4}$ for each spin structure.

 In order to get the spacetime contribution to the
 boundary state of the D-brane moving with constant velocity V,
 let us  consider the  boost \cite{divec} $|B_v> = e^{-i v^j J^0_j} |B>$,
 where $V=tanh ~v$, ($v=|v^{j}|$), is the velocity
  and $J^{\mu \nu}$ is the Lorentz generator. 
 
The full amplitude is a product of the amplitudes for the bosonic and
fermionic coordinates. We first consider the bosonic coordinates.
With the standard commutation relations for the
bosonic oscillators 
 $ [ a_n^\nu,a_{-m}^{\nu}] =[\tilde{a}_n^\nu,\tilde{a}_{-m}^{\nu}]=
\eta^{\mu\nu} \delta_{nm}$, the oscillators, ($\alpha_n, \beta_n$), for
the $X^A$ and $X^B$
fields, respectively, are now defined as $\alpha_{n}= a^{0}_{n}+a^{1}_{n},
\beta_{n}=a^{0}_{n}-a^{1}_{n}$, etc., with the commutation relations
$[\alpha_n,\beta_{-m}]= [\tilde{\alpha}_n,\tilde{\beta}_{-m}]= 
-2\delta_{nm}$, and the other commutators being zero. 

The Neumann boundary conditions for the time and
Dirichlet for the space coordinates translate into

\beq
(\alpha_n+\tilde{\beta}_{-n})|B>=0,\quad
(\beta_n+\tilde{\alpha}_{-n})|B>=0\,.
\eeq
Here $|B>$ is the unboosted bosonic spacetime part of the boundary
state. Under a Lorentz boost in the $1$ direction the
oscillators transform as
\beq
\alpha_n\rightarrow e^{-v}\alpha_n\quad
\beta_n\rightarrow e^{v}\beta_n
\eeq
and similarly for the $\tilde{\alpha}_n, \tilde{\beta}_n$. The bosonic
spacetime part of the boundary state of the moving brane is
then
 \beq
 |B_v>=  exp\sum_{n>0}\{{1\over 2}(e^{-2v}\alpha_{-n} 
\tilde{\alpha}_{-n} + e^{2v} \beta_{-n}
 \tilde{\beta}_{-n}) +a^T_{-n}.\tilde{a}^T_{-n}\}|0>\,,
 \eeq
\noindent where $a^{T}_{n}$ denote the oscillators of the
 directions orthogonal to the motion of the brane.
The momentum content of the boosted state will be $q^0= sinh(v)q^{1}$, 
$\vec{q}=(cosh(v)q^1, q_{\perp})$ Therefore from momentum conservation in
eq. (\ref{ampl}) we will get
$ q^{1} = k^{1} = 0$, $q_{\perp}=k_{\perp}$ and thus we get the amplitude at
fixed
impact parameter $Y_{\perp}$ as
\beq
{\cal A}=\int d^{3}k_{\perp}e^{ik_{\perp}Y_{\perp}}\int_0^{\infty}dl
M(l,k_{\perp})\,.
\eeq
In the following we write $M=Z_B Z_F$ where $Z_{B,F}$ are the bosonic,
fermionic contributions.
We start by computing the matrix element representing the
bosonic spacetime coordinates contribution to the amplitude
 \beq
 Z_B = <B_v| e^{-lH} |B> \label{matel}
 \eeq
 where  $H$ is the usual closed string Hamiltonian.

The oscillator part of the bosonic spacetime contribution is computed to be
\cite{divec}
\beq \prod_{n=1}^{\infty}\frac{1}{(1-e^{-2v}e^{-4\pi ln})
(1-e^{2v}e^{-4\pi ln})} 
= {2f(q^2)q^{1/4}i {sinh v} \over\vartheta_1({iv \over {\pi}}|2il)}\,, 
\eeq 
where $f(q^2)=\prod_{n=1}^{\infty}(1-q^{2n})$
with $q=e^{-2\pi l}$.

Let us now introduce the standard $Z_3$ orbifold \cite{min}, that is
compactifying the $\mu=4,5,6,7,8,9$ coordinates on a 6-torus and
identifying points which are equivalent under $g_a=e^{2\pi iz_a}$
rotations
on pairs of them, with $z_{4,5}=z_{6,7}=\pm 1/3$ and 
$z_{8,9}= -z_{4,5}-z_{6,7}$ for the pairs
$X^{4,5}=X^4+iX^5$, $X^{6,7}=X^6+iX^7$, $X^{8,9}=X^8+iX^9$ respectively.
 The request of the same 
Neumann or Dirichlet b.c.s for  both members of a pair is, 
 with $\beta^{4,5}_n =a^4_n+ia^5_n$,  $\beta^{4,5*}_n =a^4_n-ia^5_n$ etc.
\beq
(\beta^a_n \pm\tilde\beta^{a}_{-n} )|B>=0~~~~
(\beta^{a*}_n \pm\tilde\beta^{a*}_{-n} )|B>=0
\eeq
and the corresponding boundary state is
\beq
|B>=\prod_a exp{\mp\frac{1}{2}\sum_{n\geq 1}
      (\beta^a_{-n}\tilde\beta^{a*}_{-n}
      +\beta^{a*}_{-n}\tilde\beta^{a}_{-n} )}|0>\,.
     \label{torus}
\eeq
This is the same as for the torus compactification, except that 
in the orbifold case there could be a twist in the $\sigma$-direction
giving noninteger moding, which we discuss in a while.
However the presence of the
orbifold opens new possibilities for BPS states. 
In fact, let us consider D3-branes in TypeIIB theory 
with Neumann b.c.s for
$X^i$ and Dirichlet b.c.s for $X^{i+1}$ ($i=4,6,8$) at both ends
$\tau =0,l$. That is
\beq
(\beta^a_n +\tilde\beta^{a*}_{-n} )|B>=0
~~~~~(\beta^{a*}_n +\tilde\beta^a_{-n} )|B>=0
\eeq
and the corresponding boundary state is
\beq
|B>=\prod_a exp{-\frac{1}{2}\sum_{n\geq 1}
      (\beta^a_{-n}\tilde\beta^a_{-n}
      +\beta^{a*}_{-n}\tilde\beta^{a*}_{-n} )}|0>.
   \label{mix}
\eeq
The physical boundary state, which is required to be $Z_3$ invariant, is
\beq
|B_{phys}>=\frac{1}{3} (|B,1>+|B,g>+|B,g^2>) \label{inv}
\eeq
where we have introduced the "twisted boundary state", 
$(g^a\beta^a_n +g^{a*}\tilde\beta^{a*}_{-n} )|B,g>=0$:
\beq
|B,g>=\prod_a exp{-\frac{1}{2}\sum_{n\geq 1}
      ( (g^a)^2\beta^a_{-n}\tilde\beta^a_{-n}
      +(g^{a*})^2\beta^{a*}_{-n}\tilde\beta^{a*}_{-n} )}|0>\,.
       \label{fund}
\eeq
Since $g^2$ is generically an element of $Z_3$ we will write, in the
following,  $g\beta\tilde\beta$ for $g^2\beta\tilde\beta$.
One gets for a pair of coordinates 
($g_{a}^{*}g_{a}^{\prime}=e^{2\pi iz_{a}}$)
\beq
<B_a,g_{a}|e^{-lH}|B_a,g^{\prime}_{a}>=
\prod_{n=1}^{\infty}\left|\frac{1}{1-g_{a}^{*}g_{a}^{\prime}e^{-4\pi
ln}}\right|^{2}
=\frac{2f(q^{2})q^{1/4}sin(\pi z_{a})}{\vartheta_{1}(z_{a}|2il)}
\,.
\eeq

 Taking now into account all the contributions from the compactified
directions as well as the spacetime sector and the normal ordering
term from the Hamiltonian $(q^{-2/3})$, it is seen that 
the oscillator part of the bosonic amplitude is 

\beq
Z(g,g^{\prime})_B = i\left[2f(q^2)\right]^4 {q^{1/3}sinh(v) \over
{\vartheta_1({iv\over{\pi}}|2il)}}
\prod_{a}{{sin(\pi z_{a})}\over {\vartheta_1(z_a|2il)}}\,.
\label{bospart}
\eeq

On the orbifold $\sigma$-twisted sectors are also possible. We will be  
concerned with this sector only in the case when the branes are
on an orbifold fixed point, since only then the twisted closed string
is shrinkable to zero. Thus we consider this sector only for Dirichlet
b.c.s on every compact coordinate 
(thus for D0-branes in TypeIIA) and the
corresponding boundary state is the one of eq. (\ref{torus}).
In this sector
the pairs of fields in the compactified directions may be diagonalized
\cite{min} such that

\beq
X^{a,b}(\sigma+1)=e^{2\pi i z_a}X^{a,b}(\sigma),\quad
X^{*a,b}(\sigma+1)=e^{-2\pi i z_a}X^{*a,b}(\sigma)\,,
\eeq

\noindent with $(a,b)=(4,5), (6,7),(8,9)$. This leads to fractional moding
of the oscillators.
The oscillator part of the bosonic amplitude for a pair of cordinates
$X^{a,b}$ becomes \beq
<B_{a}|e^{-lH}|B_{a}>
=\prod_{n=1}^{\infty}(1-e^{-4\pi l(n-\frac{1}{3})})^{-1}
(1-e^{-4\pi l(n-\frac{2}{3})})^{-1}\,.
\eeq
Combining with the spacetime part 
and converting to Jacobi theta functions we get the full bosonic
amplitude to be
\beq
Z_B(\sigma -twisted)= 2\left[f(q^2)\right]^4{sinh(v) \over
{\vartheta_1({iv\over{\pi}}|2il)}}
\vartheta_1(-2il/3|2il)^{-3}\,.
\eeq

Now we consider the fermionic modes' contribution. Again here we combine
the fermionic coordinates $\psi^0$
and $\psi^1$ into a pair of coordinates $\psi^A=\psi^0+\psi^1$ and
$\psi^B=\psi^0-\psi^1$. The oscillators satisfy the
anti-commutation relations $\{\psi^A_m,\psi^B_n\}=
\{\tilde{\psi}^A_m,\tilde{\psi}^B_n\}=-2\delta_{mn}$ with appropriate
half-integer or integer moding for the NS-NS and RR cases. The other
directions
are combined into complex pairs $(2,3)$,$(4,5)$,$(6,7)$, $(8,9)$ as
before. With the appropriate normalisation the spacetime
modes' b.c.s for
the D-brane at rest are given by \cite{pol2} (Neumann for time and
Dirichlet for space)
\beq
(\psi^0_n+i\eta\tilde{\psi}^0_{-n})|B,\eta>=0,\quad
(\psi^i_n-i\eta\tilde{\psi}^i_{-n})|B,\eta>=0\,.
\eeq

\noindent Here $\eta=\pm 1$ has been introduced to deal with the GSO
projection. For the longitudinal coordinates these can be rewritten as
$(\psi^A_n+i\eta\tilde{\psi}^B_{-n})|B>_{F}=0\,\,\,,
(\psi^B_n+i\eta\tilde{\psi}^A_{-n})|B>_{F}=0$

Now to construct the moving boundary state we note that under the
boost in the $X^1$ direction the fields $\psi^A$ and $\psi^B$
transform like the bosonic coordinates $\psi^{A}\rightarrow
e^{-v}\psi^{A}\,\,\,,
\psi^{B}\rightarrow e^{v}\psi^{B}$.
Thus the fermionic spacetime part of the boundary state of the
moving brane is found to be
\beq
|B_v,\eta>
=
exp{\sum_{m>0}\{\frac{i\eta}{2}(e^{-2v}\psi^{A}_{-m}\tilde{\psi}^{A}_{-m} 
+e^{2v}\psi^{B}_{-m}\tilde{\psi}^{B}_{-m})
-i\eta\psi^{T}_{-m}\tilde{\psi}^{T}_{-m}\}}\,. 
\eeq  
 
The zero modes of the four uncompactified coordinates $\psi^{\mu}$, with
$\mu=0,1,2,3$ for the R-R state can be identified with the $\gamma$-matrices
$\gamma^{\mu}=i\sqrt{2}\psi^{\mu}_{0}$ and
$\tilde{\gamma}^{\mu}=i\sqrt{2}\tilde{\psi}^{\mu}_{0}$, with
$\{\gamma^{\mu},\gamma^{\nu}\}=-2\eta^{\mu\nu}$, which act on a subspace
which is a direct product of two spinor spaces.  We define 
$a=(\gamma^{0}+\gamma^{1})/2$,
$a^{*}=(\gamma^{0}-\gamma^{1})/2$ and $b=(-i\gamma^{2}+\gamma^{3})/2$,
$b^{*}=(-i\gamma^{2}-\gamma^{3})/2$ and similarly for
$\tilde{a}$,$\tilde{b}$ and a vacuum by
$a|0>=b|0>=\tilde{a}|\tilde{0}>=\tilde{b}^{*}|\tilde{0}>=0$. 
Under the boost we have
the transformations $a\rightarrow e^{-v}a, a^{*}\rightarrow e^{v}a^{*}$ 
leading to the boosted
boundary state $|B_v,\eta>=e^{v\gamma^0\gamma^1/2}|B, v=0>$, giving 
for the RR zero mode part of the moving boundary state
\beq
|B_v,\eta>
=\frac{e^{-v}}{\sqrt{2}}e^{-i\eta(e^{2v}a^{*}\tilde{a}^{*}
-b^{*}\tilde{b})}|0>\otimes|\tilde{0}>\,.
\eeq

The GSO projected state in both the NS-NS and
RR sectors is

\beq
|{\cal B}>_{NS,R}=\frac{1}{2}\{|B,\eta>_{NS,R}-|B,-\eta>_{NS,R}\}\,.
\eeq
We now compute the matrix element
$<{\cal B}_v|e^{-lH}|{\cal B}>_{F}$ (with the appropriate Hamiltonians for
the
NS-NS and RR sectors \cite{hin}) for the fermions. Thus the spacetime part
of the fermionic amplitude turns out to be

\beq
<B_{v},\eta|e^{-lH}|B,\eta^{\prime}>=
\prod_{n>0}(1\pm e^{2v}q^{2n})(1\pm e^{-2v}q^{2n})\cdot
Z_0(\pm)\,, \label{ferpart}
\eeq
where $n$ is an integer or half integer for NS-NS or RR, 
$\eta\eta^{\prime}=\pm$ for the two possible cases of the GSO
projection and $Z_0(\pm)=1$ for NS-NS, $Z_0(+)=2cosh(v)$ and $Z_0(-)=0$
for RR.
The $\psi^{2,3}$ contribution is cancelled by the $\beta -\gamma$ ghosts,
but in the RR $\eta\eta^{\prime}=-1$ (odd spin structure) case there
remains the zero mode of
this pair giving zero.

For the compactified coordinates one again combines the fermions into pairs
$\chi_{n}^{4,5}=\psi_{n}^{4}+i\psi_{n}^{5}$, etc.
Let us describe the fermionic part of the previously introduced
twisted boundary state (\ref{fund}) (mixed Neumann/Dirichlet b.c.s for each
pair). The condition for the twisted boundary state is now
\beq
(g^a\chi_{n}^{a}
+ig^{a*}\eta\tilde{\chi}_{-n}^{a*})|B,g_{a},\eta>=0\,,
\label{twf}
\eeq
for each pair of coordinates and the state satisfying this condition is
(for generic $a$ and $g$)
\beq
|B,g,\eta>=exp\left\{\frac{i\eta}{2}\sum_{n>0}(g\chi_{-n}\tilde{\chi}_{-n}
+g^{*}\chi_{-n}^{*}\tilde{\chi}_{-n}^{*}\right\}|0>\,.
\eeq
 Using
these states we find for a pair of fermionic
coordinates in the compactified space, $\sigma$-untwisted closed string
sector,  that

\beq
<B,g,\eta|e^{-lH}|B,g^{\prime}\eta^{\prime}>=
\prod_{n>0}\left|1\pm e^{2\pi iz_{a}}q^{2n}\right|^{2}\cdot Z_0^c(\pm)\,.
\eeq

\noindent Now $Z_0^c(\pm)=1$ for the NS-NS sector, $Z_0^c(+)=2cos\pi z_a$
and 
$Z_0^c(-)=2isin\pi z_a$ for the RR sector.

Putting everything together, taking into account all the
compactified directions as well as the spacetime contribution and the
normal ordering terms for both the NS-NS sector and the RR sector we find,
for the twisted boundary state ($\sigma$-untwisted sector), that the
fermionic amplitude is, in terms of Jacobi theta functions,

\beqa
&&<{\cal B}_v,g|e^{lH}|{\cal
B},g^{\prime}>_{F}=\frac{1}{q^{1/3}f(q^2)^4}\times\nn\\
&&\left\{\vartheta_2(\frac{iv}{\pi}|2il)
\prod_{a}\vartheta_2(z_a|2il)-\vartheta_3(\frac{iv}{\pi}|2il)
\prod_{a}\vartheta_3(z_a|2il)+\vartheta_4(\frac{iv}{\pi}|2il)
\prod_{a}\vartheta_4(z_a|2il)\right\}\,.\label{untwist}
\eeqa

\noindent In making the algebraic sum of the RR and NS-NS sectors, 
we take the same
signs which hold for the partition function on the torus. The first term
is the contribution from the RR sector (even spin structure) whereas the
second and the third
are from the two NS-NS GSO projections. Observe that the compactified
parts of eqs. (\ref{bospart}) and (\ref{untwist}) agree with the 
$\vartheta$-function expressions on the torus found in ref. \cite{min}.
This is consistent with viewing the cylinder as half of a torus, and also
the $<\psi\psi >$ correlators computed with the b.c. (\ref{twf}) 
would agree with those on the torus with Minahan's twist \cite{min} in the
$\tau$-direction.   
Since the $Z_3$ invariant, physical boundary state is given by the
linear combination of eq. (\ref{inv}), one has still to sum 
the product of the bosonic (\ref{bospart}) and fermionic (\ref{untwist})
amplitudes  over
the three possibilities for $g$ and $g^{\prime}$ (actually only three
possibilities for $g^{-1}g{\prime}$ are distinct). 

 On using the
Riemann identity it is easy to see that the amplitude (\ref{untwist}) 
behaves for small
velocities as $V^{2}$, if $g\neq g^{\prime}$ and as $V^4$ if 
$g= g^{\prime}$.  
In the limit $l\rightarrow\infty$ 
(where the bosonic part (\ref{bospart}) is $z_a$ independent)
the leading behaviour of this amplitude is proportional to  

\beq
\{4cosh(v)\prod_{a}cos(\pi z_{a}) 
-(cosh2v + \sum_{a}cos(2\pi z_{a})\}\,.\label{unt}
\eeq
We observe here that the first term with $cosh(v)$ is the contribution
of the RR vector whereas the rest are the contributions from NS-NS exchange.
We note that the amplitude (\ref{untwist}) vanishes at $v=0$ for all the
twists $(1,g,g^{2})$ individually.

Just as for the bosonic amplitude, when the position of the brane is on
the fixed point of the
orbifold here  we also have to include the closed string $\sigma$-twisted
sectors. Here too the oscillator moding is modified from the usual
integer and half-integer in the RR and NS-NS sectors \cite{hin}. 
By grouping the coordinates into pairs we find for each pair of
compactified coordinates in the NS-NS sector that
\beq
<B,\eta|e^{-lH}|B,\eta^{\prime}>_{NS}=
\prod_{n=1}^{\infty}\left[1\pm e^{-4\pi l(n-\frac{5}{6})}]
[1\pm e^{-4\pi l(n-\frac{1}{6})}\right]\,.
\eeq
For the RR sector (in this twisted sector there are no zero modes) we
have here
\beq
<B,\eta|e^{-lH}|B,\eta^{\prime}>_{R}=
\prod_{n=1}^{\infty}\left[1\pm e^{-4\pi l(n-\frac{1}{3})}]
[1\pm e^{-4\pi l(n-\frac{2}{3})}\right]\,.
\eeq

The net result in the twisted sector is that the full fermionic amplitude,
including the spacetime sector and the appropriate normal ordering
contributions, can now be written in terms of Jacobi theta functions as
 
\beqa
&&<{\cal B}_v|e^{-lH}|{\cal
B},>_{F}=f(q^2)^{-4}\left\{\vartheta_2(\frac{iv}{\pi}|2il)
\vartheta_2(-2il/3|2il)^3\right.\nn\\
&&\left.-\vartheta_3(\frac{iv}{\pi}|2il)
\vartheta_3(-2il/3|2il)^3-\vartheta_4(\frac{iv}{\pi}|2il)
\vartheta_4(-2il/3|2il)^3\right\}\,.\label{twist}
\eeqa

\noindent Recall that in the twisted sector for the $Z_3$ orbifold there
has to be a relative positive sign between the two NS sectors because of
invariance under the modular transformation $\tau\rightarrow \tau+3$.
At low velocities this amplitude goes like $V^{2}$. 

Let us now compare the large distance interactions of the two
moving branes found from the string formalism with the
field theory results. At large distances we look for the
Feynman graphs representing the exchange of massless particles.
We can have the exchange either of scalar, or  vector or 
graviton. The scalar and the graviton give attraction  while
the vector gives repulsion, since we consider two branes of the
same nature. The net result for zero velocity is zero, since
the branes are BPS states, and this is what is obtained from
the Riemann identity in the string formalism \cite{pol1}. But when the
velocity is different from zero, the various contributions
are unbalanced. By comparing the velocity dependence with
what we get from Feynman graphs we can tell which kind of
particles are actually coupled to the branes, in various
compactification cases.

 We treat the branes as spinless particles of mass 
and charge equal to 1.
The exchange of a scalar gives then
\beq
{\cal S} = \frac{1}{k_{\perp}^2}
\eeq
where $k$ is the momentum transfer between the two branes.
In the so-called eikonal approximation in which the branes go 
straight (which is the standard setting, which we follow, for
describing the branes' interaction at nonsmall distances),
$k$ has only space components, and it is orthogonal to $\vec V$.

The vector is coupled to the current, which in the eikonal
approximation is proportional to the momentum, 
$J^{\mu}(V)\equiv(cosh(v), sinh(v))$. Note that
$J^{\mu}k_{\mu}=0$. Taking one of the branes at rest, the vector exchange
is
\beq
{\cal V}= J^{\mu}(V)J_{\mu}(0) \frac{1}{k_{\perp}^2}=
       -\frac{cosh(v)}{k_{\perp}^2}
\eeq

The graviton is coupled to the brane's energy-momentum tensor
$T^{\mu\nu}=J^{\mu}J^{\nu}$. Therefore the graviton exchange  in d-dimensions
is
\beq
{\cal G}= 2(T^{\mu\nu}(V)
        -\frac{\eta^{\mu\nu}}{d-2}T^{\rho\sigma}(V)\eta_{\rho\sigma})
       T_{\mu\nu}(0) \frac{1}{k_{\perp}^2}
     = \frac{cosh(2v)+\frac{d-4}{d-2}}{k_{\perp}^2}\,.   
\eeq
Thus we see that we can tell the nature of the
various contributions to the branes' interaction by looking at the rapidity
dependence of the $l \to \infty$ limit of the amplitude $Z_B\cdot Z_F$
(after removing the $sinh(v)$ factor in the denominator,
the eikonal amplitude being the relativistically invariant
interaction divided by that factor).

We now discuss various cases.

1.) Toroidal compactification. The boundary states can be of 
the form of eqs. (\ref{torus}) or (\ref{mix}) and in any case one gets 
eqs. (\ref{untwist}) and (\ref{unt}) with $z_a=0$.
Thus in the field theory limit the amplitude is proportional to

\beq
4 cosh(v)-  cosh(2v) - 3\,.\label{tor}
\eeq  

Take in particular the case of two D0-branes.
 If we consider eq. (\ref{tor}) as the result for the ten dimensional
uncompactified case we write it as $4cosh(v)-(cosh(2v)+\frac{3}{4}) 
- \frac{9}{4}$.
The first term is the contribution from the exchange
of the RR vector whereas the second term is from the exchange of the
graviton in 10 dimensions (as is seen from the field theory calculation
above) and the last is the scalar exchange
contribution. We have normalised to the graviton exchange.
Our D0-brane in 10 dimensions
is like the classical 0-brane solution \cite{stelle} of the bosonic part
of the Type IIA 10 dimensional supergravity action.

If we now view eq.(\ref{tor}) from the point of view of
toroidal compactification, with no SUSY breaking, in four dimensions the 
graviton exchange gives the contribution $cosh(2v)$.  Our toroidal
compactification
corresponds to Stelle's vertical reduction of the 10 dimensional 0-brane
solution to four dimensions \cite {stelle}. In this case the relations between 
the masses, the electric charge and the scalar coupling of the 0-brane are 
$Q^{2}=4M^{2}$ and $a^{2}=3M^{2}$, precisely the relations that we have
obtained.
Further, in both these cases,
uncompactified and toroidally compactified, the force between the branes goes
to zero like $V^4$ as expected as there is no supersymmetry breaking.
Following the work of Pollard \cite{pollard} we also see that the
D0-branes in this case are  
extremal Dobiasch-Maison blackholes \cite{dm}. This extremal blackhole has zero
horizon radius and zero horizon area. 

2.) Orbifold compactification, $\sigma$-untwisted sector. For the
interaction of
two branes with the same Neumann or Dirichlet b.c.s for both members of
pairs of compactified coordinates (thus for TypeIIA)  we have the 
same result as for toroidal compactification. 
If instead we take the two branes to be in the mixed Neumann/Dirichlet
configuration corresponding to the boundary state of eq. (\ref{inv})
(thus for D3-branes in TypeIIB), we find, by summing eq. (\ref{unt}) 
over the allowed $z_a$'s, that
at large brane separation the amplitude is proportional to

\beq
cosh(v)-cosh(2v)\,.\label{orbunt}
\eeq

\noindent Thus, these two branes interact through the exchange
of the RR vector and the universal graviton with no scalar exchange. In 
terms of the N = 2 SUSY theory these systems couple only to
the graviton and its N = 2 partner, the graviphoton. In this case the
amplitude behaves like $V^2$ for small velocities. From the pattern
of cancellation \cite{pollard} we see that these branes correspond to
classical extremal Reissner-Nordstr\"{o}m blackholes. 

3.) Orbifold compactification, $\sigma$-twisted sector. In this case the 
$l\to\infty$ limit of eq.(\ref{twist}) gives
 
\beq
cosh(v)-1 \label{orbtw}
\eeq

When the branes are on the fixed point (Dirichlet b.c.s for every
compactified coordinate, TypeIIA) 
 they interact through the interchange of all
three fields, because we have contributions from both untwisted
(giving in this case the toroidal result) and twisted sectors. 
Here we have extra vectors as well as scalars 
and the force between the branes
falls off as $V^2$.

It is now an easy exercise to repeat the above calculation for the
$T_4/Z_2\times T_2$ orbifold which reduces the supersymmetry to N = 4
rather than N = 2. Here in the $\sigma$-untwisted sector, 
we can also construct the mixed Neumann-Dirichlet twisted
boundary state for the first two coordinate pairs
(in this case it could be done for both TypeIIA and TypeIIB). 
However since in this case the twists 
on the two pairs of coordinates $X^4,X^5$ and $X^6,X^7$ 
are $g_a=exp(2\pi iz_a)$ with
$z_{4,5}=-z_{6,7}=1/2$, and thus $g^2\beta\tilde\beta=\beta\tilde\beta$, 
then in the
notation of  eqs. (\ref{inv}),(\ref{fund})  
$|B,g>=|B,1>$ and $|B_{phys}>=|B,1>$.
Therefore the interaction between two parallel branes 
in this case 
behaves like the toroidal compactification, the force falling off like $V^4$
 \cite{ooguri}.

For the $\sigma$-twisted sector the fermionic partition function is
proportional to  \beqa
&&\left\{\vartheta_2(\frac{iv}{\pi}|2il)\vartheta_2(0|2il)
\vartheta_2(-il|2il)^2\right.\nn\\
&&\left.-\vartheta_3(\frac{iv}{\pi}|2il)\vartheta_3(0|2il)
\vartheta_3(-il|2il)^2
-\vartheta_4(\frac{iv}{\pi}|2il)\vartheta_4(0|2il)
\vartheta_4(-il|2il)^2\right\}\,.\label{twistk3}
\eeqa

\noindent Again in the twisted sector for large separation we find that
the amplitude is proportional to $ cosh(v) - 1$ and the
force falls off as $V^2$.  
  
{\bf Acknowledgements}.
The authors would like to thank E. Gava and K.S. Narain for very valuable
discussions. C.A.S. also thanks F. Morales and M. Serone for useful
exchange of ideas. C.N. would like to thank ICTP for
hospitality during the completion of this work, which was done in the
frame of the Associate Membership Programme of the ICTP. R.I. and C.A.S. 
acknowledge partial support from EEC contract ERBFMRXCT96-0045.


\begin{thebibliography}{99}
 \bibitem{bachas} C. Bachas, Phys. Lett. {\bf 374B} (1996) 37;\\
 C. Bachas, {\it (Half) a lecture on D-branes}, Workshop on
 Gauge Theories, Applied Supersymmetry and Quantum Gravity, Imperial
 College, London (July 1996).
 \bibitem{lifschytz} G. Lifschytz, {\it ``Comparing D-branes to
 Black-branes''},
 hep-th/9604156
 \bibitem{porrati} C. Bachas and M. Porrati, Phys. Lett. {\bf B296} (1992) 77 
\bibitem{kp} D. Kabat and P. Pouliot, {\it ``A Comment on Zero-Brane
 Quantum Mechanics''}, hep-th/9603127;
 \\ U. H. Danielsson, G. Ferretti and
 B. Sundborg, {\it ``D-Particle Dynamics and Bound States''}, hep-th/9603081
 \bibitem{shenker} S. H. Shenker, {\it ``Another length scale in String
 Theory?''}, hep-th/9509132;\\
 M. Douglas, D. Kabat, P. Pouliot and S. Shenker,
 {\it ``D-branes and Short Distances in String Theory''}, 
 hep-th/9608024
\bibitem{mth} E. Witten, Nucl. Phys. {\bf B443} (1995) 85\\
J. Schwarz, Phys. Lett. {\bf B360} (1995) 13; {\bf B367} (1996) 97
\bibitem{bh} A. Strominger and C. Vafa, {\it "Microscopic origin of the
Bekenstein-Hawking entropy"}, hep-th/9601029;\\
C.G. Callan,Jr. and J.M. Maldacena, {\it "D-brane approach to Black Hole
Quantum Mechanics"}, hep-th/9602043;\\
G. Horowitz and A. Strominger, {\it "Counting States of Near-Extremal
Black Holes"}, hep-th/9602051;\\
C.V. Johnson, R.R. Khuri and R.C. Myers, {\it "Entropy of 4D Extremal
Black Holes"}, hep-th/9603061
\bibitem{pol2} J. Polchinski and Y. Cai, Nucl. Phys. {\bf B296} (1988) 91
 \bibitem{min}J. A. Minahan, Nucl. Phys. {\bf B298} (1988) 36
 \bibitem{hin} F. Hussain, R. Iengo and C. N\'u\~nez, {\it Axion
 production from gravitons off interacting 0-branes}, hep-th/9701143, IC/97/1,
 SISSAREF-3/97/EP, to appear in Nucl. Phys. {\bf B}
 \bibitem{divec} M. Bill\'o, P. Di Vecchia and D. Cangemi, {\it 
 ``Boundary
 states for moving D-branes''}, hep-th/9701190, NBI-HE-97-05, NORDITA 97/7P
\bibitem{pol1} J. Polchinski, Phys. Rev Lett. {\bf 75} (1995) 4724 
\bibitem{stelle} K.S. Stelle, {\it Lectures on supergravity p-branes},
ICTP Summer School in High Energy Physics and Cosmology, Trieste,
June 10-26, 1996. hep-th/9701088
 \bibitem{pollard} D. Pollard, J. Phys.{\bf A 16} (1983) 565 
\bibitem{dm} P. Dobaisch and D. Maison, GRG {\bf 14} (1982) 393 
\bibitem{ooguri} M.R. Douglas, H. Ooguri and S.H. Shenker,
{\it "Issues in M(atrix) Theory Compactification"}, hep-th/9702203
 \end{thebibliography}
 \end{document}